\documentclass[12pt]{article}

\usepackage{amsmath,amsfonts,amssymb}

\begin{document}

\begin{titlepage}

\begin{center}

\begin{center}
{\Large{ \bf Integrability of D1-brane on Group Manifold}}
\end{center}

\vskip 1cm

{\large Josef Kluso\v{n}$^{}$\footnote{E-mail: {\tt
klu@physics.muni.cz}} }

\vskip 0.8cm

{\it Department of
Theoretical Physics and Astrophysics\\
Faculty of Science, Masaryk University\\
Kotl\'{a}\v{r}sk\'{a} 2, 611 37, Brno\\
Czech Republic\\
[10mm]}

\vskip 0.8cm

\end{center}

\begin{abstract}
This paper  is devoted to the analysis of the integrability of
D1-brane  on  group manifold. We consider D1-brane as principal
chiral model, determine corresponding equations of motions and find
Lax connection. Then we calculate the Poisson brackets of Lax
connection and we find that it has similar structure as in case of
principal chiral model. As the second example  we  consider more
general background with non-zero NS-NS two form.  We again show that
D1-brane theory is integrable on this background and determine
Poisson brackets of Lax connection.
\end{abstract}

\bigskip

\end{titlepage}

\newpage

\newcommand{\mT}{\mathcal{T}}
\def\tr{\mathrm{Tr}}
\def\tJ{\tilde{J}}
\def\str{\mathrm{Str}}
\newcommand{\tL}{\tilde{L}}
\def\Pf{\mathrm{Pf}}
\def\I{\mathbf{i}}
\def\IT{\I_{\Phi,\Phi',T}}
\def \cit{\IT^{\dag}}
\def \cdt{\overline{\tilde{D}T}}
\def \dt{\tilde{D}T}
\def\bra #1{\left<#1\right|}
\def\ket #1{\left|#1\right>}
\def\vac #1{\left<\left<#1\right>\right>}
\def\pb  #1{\left\{#1\right\}}
\def \uw #1{(w^{#1})}
\def \dw #1{(w_{#1})}
\newcommand{\bK}{\mathbf{K}}
\newcommand{\thw}{\tilde{\hat{w}}}
\newcommand{\bA}{{\bf A}}
\newcommand{\bd}{{\bf d}}
\newcommand{\bD}{{\bf D}}
\newcommand{\bF}{{\bf F}}
\newcommand{\bN}{{\bf N}}
\newcommand{\hp}{\hat{p}}
\newcommand{\hq}{\hat{q}}
\newcommand{\hF}{\hat{F}}
\newcommand{\hG}{\hat{G}}
\newcommand{\hH}{\hat{H}}
\newcommand{\hU}{\hat{U}}
\newcommand{\mH}{\mathcal{H}}
\newcommand{\mG}{\mathcal{G}}
\newcommand{\mA}{\mathcal{A}}
\newcommand{\mD}{\mathcal{D}}
\newcommand{\tpr}{t^{\prime}}
\newcommand{\bzg}{\overline{\zg}}
\newcommand{\of}{\overline{f}}
\newcommand{\ow}{\overline{w}}
\newcommand{\htheta}{\hat{\theta}}
\newcommand{\opartial}{\overline{\partial}}
\newcommand{\hd}{\hat{d}}
\newcommand{\halpha}{\hat{\alpha}}
\newcommand{\hbeta}{\hat{\beta}}
\newcommand{\hdelta}{\hat{\delta}}
\newcommand{\hgamma}{\hat{\gamma}}
\newcommand{\hlambda}{\hat{\lambda}}
\newcommand{\hw}{\hat{w}}
\newcommand{\hN}{\hat{N}}
\newcommand{\onabla}{\overline{\nabla}}
\newcommand{\hmu}{\hat{\mu}}
\newcommand{\hnu}{\hat{\nu}}
\newcommand{\ha}{\hat{a}}
\newcommand{\hb}{\hat{b}}
\newcommand{\hc}{\hat{c}}
\newcommand{\com}[1]{\left[#1\right]}
\newcommand{\oz}{\overline{z}}
\newcommand{\oJ}{\overline{J}}
\newcommand{\mL}{\mathcal{L}}
\newcommand{\oh}{\overline{h}}
\newcommand{\oT}{\overline{T}}
\newcommand{\oepsilon}{\overline{\epsilon}}
\newcommand{\tP}{\tilde{P}}
\newcommand{\hP}{\hat{P}}
\newcommand{\talpha}{\tilde{\alpha}}
\newcommand{\uc}{\underline{c}}
\newcommand{\ud}{\underline{d}}
\newcommand{\ue}{\underline{e}}
\newcommand{\uf}{\underline{f}}
\newcommand{\hpi}{\hat{\pi}}
\newcommand{\oZ}{\overline{Z}}
\newcommand{\tg}{\tilde{g}}
\newcommand{\tK}{\tilde{K}}
\newcommand{\tj}{\tilde{j}}
\newcommand{\tG}{\tilde{G}}
\newcommand{\hg}{\hat{g}}
\newcommand{\htG}{\hat{\tilde{G}}}
\newcommand{\hX}{\hat{X}}
\newcommand{\hY}{\hat{Y}}
\newcommand{\bDi}{\left(\bD^{-1}\right)}
\newcommand{\hthteta}{\hat{\theta}}
\newcommand{\hB}{\hat{B}}
\newcommand{\tlambda}{\tilde{\lambda}}
\newcommand{\thlambda}{\tilde{\hat{\lambda}}}
\newcommand{\tw}{\tilde{w}}
\newcommand{\hJ}{\hat{J}}
\newcommand{\tPsi}{\tilde{\Psi}}
\newcommand{\cP}{{\cal P}}
\newcommand{\tphi}{\tilde{\phi}}
\newcommand{\tOmega}{\tilde{\Omega}}
\newcommand{\homega}{\hat{\omega}}
\newcommand{\hupsilon}{\hat{\upsilon}}
\newcommand{\hUpsilon}{\hat{\Upsilon}}
\newcommand{\hOmega}{\hat{\Omega}}
\newcommand{\bJ}{\mathbf{J}}
\newcommand{\olambda}{\overline{\lambda}}
\newcommand{\uhlambda}{\underline{\hlambda}}
\newcommand{\uhw}{\underline{\hw}}
\newcommand{\tC}{\tilde{C}}
\def \lhw #1{(\hw^{#1})}
\def \dhw #1{(\hw_{#1})}
\newcommand{\bG}{\mathbf{G}}
\newcommand{\bhG}{\hat{\bG}}
\newcommand{\bH}{\mathbf{H}}
\newcommand{\bE}{\mathbf{E}}
\newcommand{\mJ}{\mathcal{J}}
\newcommand{\mY}{\mathcal{Y}}
\newcommand{\mZ}{\mathcal{Z}}
\newcommand{\hj}{\hat{j}}
\newcommand{\bAi}{\left(\bA^{-1}\right)}
\newcommand{\hpartial}{\hat{\partial}}
\newcommand{\hD}{\hat{D}}
\newcommand{\mC}{\mathcal{C}}
\newcommand{\omC}{\overline{\mC}}
\newcommand{\mP}{\mathcal{P}}
\newcommand{\omP}{\overline{\mP}}
\newcommand{\tLambda}{\tilde{\Lambda}}
\newcommand{\tPhi}{\tilde{\Phi}}
\newcommand{\tD}{\tilde{D}}
\newcommand{\tgamma}{\tilde{\gamma}}
\section{Introduction and Summary}
One of the greatest achievements in string theory is the discovery
of the integrability of the classical string sigma model on
$AdS_5\times S^5$ background \cite{Bena:2003wd} \footnote{For recent
review of integrability in the context of string theory, see
\cite{Sfondrini:2014via,Puletti:2010ge,vanTongeren:2013gva,Arutyunov:2009ga}}.
The integrability of the string is based on the existence of the Lax
connection that leads to the existence of the infinite tower of
conserved charges. However as was stressed in \cite{Dorey:2006mx}
the integrability means that we have infinite tower of conserved
charges that are also in involution which means that they Poisson
commute with each other.  On the other hand even if the theory
possesses Lax formulation, there is a well known problem in
determining the Poisson brackets of the conserved charges due to the
existence of non-ultra local terms in these Poisson brackets. One
possibility how to resolve this problem is in the prescription of
regularizing of these problematic brackets that was known from the
work by Maillet \cite{Maillet:1985ek,Maillet:1985ec}, for
alternative possibility, see recent work
\cite{Delduc:2012qb,Delduc:2012vq}.

As is well known string theories contain another extended objects,
as for example Dp-branes, NS5-branes, etc.  It is also well known
that Type IIB superstring theory is invariant under S-duality that
maps theory at week coupling to the theory at strong coupling and
where for example fundamental string is mapped to D1-brane.Then we
can  ask the question whether integrability is also preserved under
S-duality. Of course, this is very difficult problem in the full
generality due to the fact that  it is not clear whether classical
description can be applied when the coupling constant is strong. On
the other hand  we would like to see whether D1-brane, that
propagates on some group manifold, possesses integrable structure as
fundamental string does. The goal of this paper is to answer this
question. We  study  D1-brane on the background where string
propagating on given background is defined by principal chiral model
that is well known to be integrable. On the other hand the dynamics
of D1-brane is governed by Dirac-Born-Infeld action with presence of
the gauge field so that the integrability of given theory has to be
checked. We explicitly construct corresponding Lax pair in case of
D1-brane on the group manifold  and show that obeys the flatness
conditions on condition when all fields obey the equations of
motion. We further proceed to the Hamiltonian formalism and
calculate the Poisson bracket between spatial components of Lax
connection and we find that it has the same form as in case of
principal chiral model with exceptions that coefficients  depend on
the momentum conjugate to the spatial component of the gauge field
and we also find  that the resulting Poisson bracket contains the
constraint that corresponds to the gauge invariance of given theory.

As the next step we proceed to the analysis of the integrability of
D1-brane on the manifold that could be described as
Wess-Zummino-Witten model which means that there is non-trivial
$b_{NS}$ field.  We  determine corresponding Hamiltonian for
D1-brane and determine its constraint structure. Then we construct
Lax connection that  obeys the flatness condition when all fields
are on shell. Note that the parameters in the Lax connection, that
were determined from the requirement of its flatness, depends on the
constant (on-shell) value of the momenta conjugate to the spatial
component of the gauge field. When we proceed to the Hamiltonian
formalism  we perform the of-shell extension of this expression.
Then we calculate the Poisson bracket between spatial components of
the Lax connections and we find that it takes the same form as in
case of the WZW model with the  exception that now the parameters
depend on the momenta conjugate to the spatial component of the
gauge field and also it is proportional to the secondary constraint
that forces this momentum to be spatial independent.

In summary, we find that D1-brane possesses the same integrability
as the fundamental string on given background. The fact that
D1-brane is integrable could be considered as the first step in the
analysis of the more general configurations of D1-branes on group
manifold. More explicitly,  it would be very interesting to discuss
non-abelian DBI action for collection of $N$ D1-branes on given
background \cite{Myers:1999ps}. We can expect that when we have
collection of $N$ D1-branes that are far away that the resulting
theory should be integrable since it effectively reduces to the
collections of $N$ abelian D1-brane actions. However it would be
nice to see what happens when we move these branes closer. In this
case we can expect that the integrability is lost. We hope to return
to this problem in future.

The organization of this paper is as follows. In the next section
(\ref{second}) we consider D1-brane as principal chiral model and
determine corresponding Lax connection. In section (\ref{third}) we
calculate the Poisson brackets between spatial components of Lax
connection. In section (\ref{fourth}) we consider D1-brane as WZW
model. Finally in section (\ref{fifth}) we determine Hamiltonian
formalism for given theory and calculate the Poisson bracket of
spatial components of Lax connection.
\section{D1-brane on Group Manifold }\label{second}
Our goal is to show that D1-brane on the group manifold defines an
integrable theory.
 Let us consider
following D1-brane action
\begin{equation}
S=-T_{D1}\int d\sigma d\tau  \sqrt{-\det \bA} \ ,
\end{equation}
where
\begin{eqnarray}
& &\bA_{\alpha\beta}=g_{\alpha\beta}+2\pi \alpha'F_{\alpha\beta} \ , \nonumber
\\
& & g_{\alpha\beta}= g_{MN}\partial_\alpha x^M
\partial_\beta x^N \ , \quad  F_{\alpha\beta}=\partial_\alpha A_\beta-
\partial_\beta A_\alpha \ , \quad
 x^\alpha\equiv (\tau,\sigma) \ , \nonumber \\
\end{eqnarray}
and where D1-brane tension is equal to
\begin{equation}
T_{D1}=\frac{1}{g_s (2\pi \alpha')} \ ,
\end{equation}
where $g_s$ is constant string coupling constant.  When we consider
group manifold $G$ we presume that
 $g_{MN}$ has the form
\begin{equation}
g_{MN}=E_M^{\ A}E_N^{ \ B}K_{AB} \ ,
\end{equation}
where for the group element  $g\in G$ we have
\begin{equation}
g^{-1}dg=E_M^{ \ A}T_A dx^M \ ,
\end{equation}
 where $T_A$ is basis of Lie
Algebra $\mathcal{G}$ of the group $G$. Finally $x^M$ are
coordinates on the group manifold $G$. Now from this definition it
is clear that we can write $g_{\alpha\beta}$ as
\begin{equation}
g_{\alpha\beta}=J_\alpha^AJ_\beta^BK_{AB} \ ,
\end{equation}
 where
\begin{equation}
J_\alpha=g^{-1}\partial_\alpha g=J^A_\alpha T_A \ , J^A_\alpha=E_M^{ \ A}\partial_\alpha
x^M  \ .
\end{equation}
It is useful to rewrite the D1-brane action into the form
\begin{eqnarray}\label{D1alternative}
S
=-T_{D1} \int d^2\sigma \sqrt{-\det g-(2\pi\alpha')^2
(F_{\tau\sigma})^2} \ ,
\nonumber \\
\end{eqnarray}
where $\det
g=g_{\tau\tau}g_{\sigma\sigma}-g_{\tau\sigma}g_{\tau\sigma}$.  Now
the equation of motion for $A_\tau, A_\sigma$ that follow from
(\ref{D1alternative}) have the form
\begin{eqnarray}
& &\partial_\tau\left[\frac{2\pi\alpha'F_{\tau\sigma}} {\sqrt{-\det
g-(2\pi\alpha')^2(F_{\tau\sigma})^2}}\right]=0 \ ,
\nonumber \\
& &\partial_\sigma\left[\frac{2\pi\alpha'F_{\tau\sigma}}
{\sqrt{-\det g-(2\pi\alpha')^2(F_{\tau\sigma})^2}}\right]=0 \
\nonumber \\
\end{eqnarray}
and consequently we obtain
\begin{eqnarray}\label{defPi}
\frac{2\pi \alpha'F_{\tau\sigma}}{ \sqrt{-\det
g-(2\pi\alpha')^2(F_{\tau\sigma})^2}}=\Pi \ , \nonumber
\\
\end{eqnarray}
where $\Pi$ is a constant.
In order to derive the equation of motion for $J^A_\alpha$ let us
 consider the variation of $g$ as $\delta g=g\delta X,\ \delta
X=\delta X^A T_A$. Then we obtain
\begin{equation}
\delta J^A_\alpha =J^B_\alpha f_{BC}^{ \quad A} \delta X^C+
\partial_\alpha \delta X^A \ ,
\end{equation}
where $f_{BC}^{ \quad  A}$ are structure coefficients of Lie algebra
$\mG$ defined by $[T_A,T_B]=f_{AB}^{ \quad C}T_C$.
Then we find that  the equations of motion for the current $J^A$
take the form
\begin{equation}\label{eqJalpha}
\partial_\alpha \left[
J^A_\beta g^{\beta\alpha} \frac{\det g}{\sqrt{-\det g-
(2\pi\alpha')^2F^2_{\tau\sigma}}}\right]=0 \ .
\end{equation}
Deriving both equations of motion for $J^A_\alpha$ and $A_\alpha$ we
can now proceed to the construction of the flat current for D1-brane
on the  group manifold. Before we do it note that when we consider
equations of motion we demand that all fields are on-shell. This
fact however implies that we should  replace $F_{\tau\sigma}$ in
(\ref{eqJalpha}) with the value that follows from (\ref{defPi}) so
that (\ref{eqJalpha}) simplifies considerably
\begin{eqnarray}\label{eqJalphaPi}
\partial_\beta[J^A_\alpha g^{\alpha\beta}\sqrt{-\det
g}]=0 \ .  \nonumber \\
\end{eqnarray}
This result suggests that it is natural to consider the flat current
in the same form as in case of the fundamental string moving on the
same group manifold.  Explicitly, we consider flat current in the
form
\begin{equation}
L_\alpha=\frac{1}{1-\Lambda^2}
[J_\alpha^A-\Lambda\epsilon_{\alpha\beta} g^{\beta \omega}J_\omega^A
\sqrt{-\det g}]
\end{equation}
or in components (using
$\epsilon_{\alpha\beta}=\epsilon_{\tau\sigma}=-\epsilon_{\sigma\tau}=1$)
\begin{eqnarray}
L_\sigma=\frac{1}{1-\Lambda^2} [J_\sigma^A+\Lambda
g^{\tau\alpha}J_\alpha^A \sqrt{-\det g}] \ , \nonumber \\
L_\tau=\frac{1}{ 1-\Lambda^2}[J_\tau^A-\Lambda
g^{\sigma\alpha}J_\alpha^A \sqrt{-\det g}] \ , \nonumber \\
\end{eqnarray}
where  $\Lambda$ is spectral parameter. Then we
 calculate
\begin{eqnarray}
\partial_\tau L_\sigma^A-\partial_\sigma L_\tau^A=
-\frac{1}{1-\Lambda^2}J^B_\tau J^C_\sigma f_{BC}^{\quad A}
\nonumber \\
\end{eqnarray}
using equations of motion for $J_\alpha^A$ and $A_\alpha$. On the
other hand we have
\begin{eqnarray}
L^B_\tau L^C_\sigma f_{BC}^{\quad A}=
\frac{1}{1-\Lambda^2}J^B_\tau J^C_\sigma f_{BC}^{\quad A} \ .
\nonumber \\
\end{eqnarray}
Collecting there results together we find
\begin{equation}
\partial_\tau L_\sigma^A-\partial_\sigma L_\tau^A+L_\tau^C
L_\sigma^D f_{CD}^{ \quad  A}=0 \ .
\end{equation}
 In other words we have
shown that the Lax connection  is flat which is  the necessary
condition for the theory to be integrable. As the next step in
the proof of the integrability we determine the
 Poisson brackets between spatial components of Lax
connection and show that it has the right form for the existence
of infinite number of conserved charges that are in involution.

\subsection{Hamiltonian Analysis and Poisson Brackets of Lax Connection}
\label{third} In this section we develop the Hamiltonian formalism
for the action (\ref{D1alternative}) and calculate the Poisson
bracket between spatial components of Lax connection. First of all
 we determine momenta conjugate to $x^M,
A_\alpha$ from (\ref{D1alternative})
\begin{eqnarray}
p_M
&=&T_{D1}\frac{g_{MN}\partial_\alpha x^N g^{\alpha\tau}\det g}
{\sqrt{
-\det g-(2\pi\alpha')^2(F_{\tau\sigma})^2}} \ , \nonumber \\
\pi^\sigma &=&-T_{D1} \frac{(2\pi\alpha')^2F_{\tau\sigma}} { \sqrt{
-\det g-(2\pi\alpha')^2(F_{\tau\sigma})^2}}
 \ , \quad
\pi^\tau\approx  0 \ . \nonumber \\
\end{eqnarray}
Using these results we find that the theory possesses two primary
constraints
\begin{eqnarray}
\mH_\tau&=&p_M g^{MN}p_N+\frac{1}{(2\pi\alpha')^2} \pi^\sigma
g_{\sigma\sigma} \pi^\sigma+T_{D1}^2\partial_\sigma x^Mg_{MN}
\partial_\sigma x^N\approx 0 \  , \nonumber \\
\mH_\sigma&=& p_M\partial_\sigma x^M \approx 0 \ . \nonumber \\
\end{eqnarray}
 On the other hand the
bare Hamiltonian has the form
\begin{eqnarray}
H_E=\int d\sigma (p_M\partial_\tau x^M+\pi^\sigma\partial_\tau
A_\sigma-L_{D1})=-\int d\sigma\pi^\sigma\partial_\sigma A_\tau \nonumber \\
\end{eqnarray}
and hence the Hamiltonian including all constraints has the form
\begin{equation}
H=\int d\sigma (\lambda_\tau\mH_\tau+\lambda_\sigma\mH_\sigma+A_\tau
\partial_\sigma \pi^\sigma+v_\tau \pi^\tau) \ ,
\end{equation}
where of course the requirement of the preservation of the
constraint $\pi^\tau\approx 0$ gives the secondary constraint
\begin{equation}
\mG=\partial_\sigma \pi^\sigma\approx 0 \ .
\end{equation}
We are not going to determine the algebra of constraints which will
be performed in the section devoted to the analysis of D1-brane on
the group manifold with non-zero NS-NS two form. We rather express
the spatial component of the flat current $L_\sigma$ as function of
the canonical variables
\begin{equation}
L^A_\sigma(\Lambda)=
\frac{1}{1-\Lambda^2} \left[E_M^{ \ A}\partial_\sigma x^M-\Lambda
\frac{K^{AB}E_{ \ B}^{
M}p_M}{\sqrt{T_{D1}^2
+\left(\frac{\pi^\sigma}
{2\pi\alpha'}\right)^2}}\right] \ .
\end{equation}
Now using the canonical Poisson brackets
\begin{equation}
\pb{x^M(\sigma),p_N(\sigma')}=\delta^M_N\delta(\sigma-\sigma') \ ,
\quad
\pb{A_\sigma(\sigma),\pi^\sigma(\sigma')}=\delta(\sigma-\sigma') \
\end{equation}
we  obtain
\begin{eqnarray}\label{pbLAB}
& &\pb{L^A_\sigma(\Lambda),L^B_\sigma(\Gamma)}
=\frac{\Lambda+\Gamma}{(1-\Lambda^2)(1-\Gamma^2)}K^{AB}
\frac{1}{\sqrt{T_{D1}^2+\left(\frac{\pi^\sigma}
{2\pi\alpha'}\right)^2}}\partial_\sigma
\delta(\sigma-\sigma')-\nonumber \\
&-&\frac{\Gamma}{(1-\Lambda^2)(1-\Gamma^2)} \frac{K^{AB}}{
(T_{D1}^2+ \left(\frac{\pi^\sigma}{2\pi\alpha'}\right)^2)^{3/2}}
\frac{\pi^\sigma}{(2\pi\alpha')^2}\mG\delta(\sigma-\sigma')+ \nonumber \\
&+&\frac{1}{(1-\Lambda^2)(1-\Gamma^2)}\frac{ (\Lambda+ \Gamma)}{
\sqrt{T_{D1}^2+ \left(\frac{\pi^\sigma} {2\pi\alpha'}\right)^2}}
E_M^{  \ D}f_{CD}^{ \quad A}K^{BC}\partial_\sigma
x^N\delta(\sigma-\sigma')+ \nonumber \\
&-&\frac{\Lambda\Gamma}{(1-\Lambda^2)(1-\Gamma^2)}
\frac{1}{T_{D1}^2+\left(\frac{\pi^\sigma}{2\pi\alpha'}\right)^2}
K^{BD}f_{DC}^{ \quad A}K^{CF} E^P_{  \ F} p_P
\delta(\sigma-\sigma') \nonumber \\
\end{eqnarray}
using
\begin{eqnarray}
& &\partial_N E_M^{\ A}-\partial_M E_N^{ \ A}+E_N^{ \ B} E_M^{ \ C}
f_{BC}^{ \quad  A}=0 \ , \nonumber \\
& & f(\sigma')\partial_\sigma
\delta(\sigma-\sigma')=f(\sigma)\partial_\sigma\delta(\sigma-\sigma')+\partial_\sigma
f(\sigma)\delta(\sigma-\sigma') \ . \nonumber \\
\end{eqnarray}
Now we demand that the expression proportional to the delta function
is equal to
\begin{eqnarray}\label{ABhelp}
-(A L^C_\sigma(\Lambda)-B L_\sigma^C(\Gamma))f_{CD}^{ \quad
A}K^{DB}\delta(\sigma-\sigma') \ . \nonumber \\
\end{eqnarray}
Comparing (\ref{ABhelp}) with (\ref{pbLAB}) we determine $A$ and $B$
as
\begin{eqnarray}
A&=&\frac{\Gamma^2}{(1-\Gamma^2)(\Gamma-\Lambda) \sqrt{T_{D1}^2+
\left(\frac{\pi^\sigma}{2\pi\alpha'}\right)^2}} \ , \nonumber \\
B&=&\frac{\Lambda^2}{(1-\Lambda^2)(\Gamma-\Lambda) \sqrt{ T_{D1}^2+
\left(\frac{\pi^\sigma}{2\pi\alpha'}\right)^2}} \ . \nonumber \\
\end{eqnarray}
In other words we find the final result
\begin{eqnarray}
& &\pb{L^A_\sigma(\Lambda),L^B_\sigma(\Gamma)}=
\frac{\Lambda+\Gamma}{(1-\Lambda^2)(1-\Gamma^2)}K^{AB}
\frac{1}{\sqrt{T_{D1}^2+\left(\frac{\pi^\sigma}
{2\pi\alpha'}\right)^2}}\partial_\sigma \delta(\sigma-\sigma')
-\nonumber \\
&-&\frac{\Gamma^2}{(1-\Gamma^2)(\Gamma-\Lambda) \sqrt{T_{D1}^2+
\left(\frac{\pi^\sigma}{2\pi\alpha'}\right)^2}} L^C_\sigma(\Lambda)
f_{CD}^{ \quad A}K^{DB}\delta(\sigma-\sigma')
-\nonumber \\
&-& \frac{\Lambda^2}{(1-\Lambda^2)(\Gamma-\Lambda) \sqrt{ T_{D1}^2+
\left(\frac{\pi^\sigma}{2\pi\alpha'}\right)^2}}
L_\sigma^C(\Gamma)f_{CD}^{ \quad A}K^{DB}\delta(\sigma-\sigma')
-\nonumber \\
&-&\frac{\Gamma}{(1-\Lambda^2)(1-\Gamma^2)} \frac{K^{AB}}{
(T_{D1}^2+ \left(\frac{\pi^\sigma}{2\pi\alpha'}\right)^2)^{3/2}}
\frac{\pi^\sigma}{(2\pi\alpha')^2}\mG\delta(\sigma-\sigma') \ .  \nonumber \\
\end{eqnarray}
We see that this Poisson bracket has similar form as in case of the
principal chiral model even if the coefficients in front of
$L_\sigma$ on the right side of the Poisson bracket depend on the
canonical variable $\pi^\sigma$ and  the right side of the Poisson
bracket contains the secondary constraint $\mG\approx 0$. On the
other hand the presence of this constraint implies that $\pi^\sigma$
does not depend on $\sigma$ on the constraint surface $\mG\approx
0$. Further, the equation of motion for $\pi^\sigma$ implies that it
does not depend on $\tau$ as well. In other words $\pi^\sigma$ is
constant on shell that physically  counts the number of fundamental
strings.
\section{D1-brane on the Background with Non-Trivial NS-NS Field }
\label{fourth}
 In this section we consider more general possibility
when D1-brane is embedded on the group manifold with non-trivial
$b_{NS}$ field. In other words we consider the background that
corresponding to the fundamental string as  WZW model which is
integrable.
The presence of this field modifies the action in the following way
\begin{equation}\label{D1Baction}
S=-T_{D1}\int d^2\sigma \sqrt{-\det g-
((2\pi\alpha')F_{\tau\sigma}+b_{\tau\sigma})^2} \ ,
\end{equation}
where
\begin{equation}
b_{\alpha\beta}\equiv b_{MN}\partial_\alpha x^M\partial_\beta x^N=
-b_{\beta\alpha} \
\end{equation}
since  $b_{MN}=-b_{NM}$. Now the equation of motion for $x^M$ has
the form
\begin{eqnarray}
 &-&\partial_\alpha\left[\frac{g_{MN}\partial_\beta x^N
g^{\beta\alpha} \det g} {\sqrt{-\det g-
((2\pi\alpha')F_{\tau\sigma}+b_{\tau\sigma})^2}}\right]
+\frac{\partial_M g_{KL}\partial_\alpha x^K \partial_\beta x^L
g^{\beta\alpha}\det g}{2 {\sqrt{-\det g-
((2\pi\alpha')F_{\tau\sigma}+b_{\tau\sigma})^2}}}
+ \nonumber \\
&+&\frac{((2\pi\alpha')F_{\tau\sigma}+b_{\tau\sigma})} {\sqrt{-\det
g-((2\pi\alpha')F_{\tau\sigma}+b_{\tau\sigma})^2}}
\partial_M b_{KL}\partial_\tau x^K\partial_\sigma x^L
-\nonumber \\
&-&\partial_\tau \left[ \frac{b_{MN}\partial_\sigma x^N
((2\pi\alpha')F_{\tau\sigma}+b_{\tau\sigma})} {\sqrt{-\det
g-((2\pi\alpha')F_{\tau\sigma}+b_{\tau\sigma})^2}}\right]
+\partial_\sigma \left[ \frac{b_{MN}\partial_\tau x^N
((2\pi\alpha')F_{\tau\sigma}+b_{\tau\sigma})} {\sqrt{-\det
g-((2\pi\alpha')F_{\tau\sigma}+b_{\tau\sigma})^2}}
\right]=0 \nonumber \\
\end{eqnarray}
while the equations of motion for $A_\tau,A_\sigma$ take the form
\begin{eqnarray}
\partial_\tau\left[\frac{(2\pi\alpha')F_{\tau\sigma}+b_{\tau\sigma}}
{\sqrt{-\det g-
((2\pi\alpha')F_{\tau\sigma}+b_{\tau\sigma})^2}}\right]=0 \ ,
\nonumber \\
\partial_\sigma\left[\frac{(2\pi\alpha')F_{\tau\sigma}+b_{\tau\sigma}}
{\sqrt{-\det g-
((2\pi\alpha')F_{\tau\sigma}+b_{\tau\sigma})^2}}\right]=0 \ .
\nonumber \\
\end{eqnarray}
These equations imply
\begin{equation}\label{soleqA}
\frac{(2\pi\alpha')F_{\tau\sigma}+b_{\tau\sigma}} {\sqrt{-\det
g-((2\pi\alpha')F_{\tau\sigma}+b_{\tau\sigma})^2}} =\Pi \ , \quad
\Pi=\mathrm{const} \ .
\end{equation}
Now with the help of the solution of the equation of motion for
$A_\alpha$ given in (\ref{soleqA}) we obtain
 that the equation of motion for
$x^M$ takes the form
\begin{eqnarray}
& &\sqrt{1-\Pi^2}\partial_\alpha[g_{MN}\partial_\beta x^N
g^{\alpha\beta} \sqrt{-\det g}]-
\frac{1}{2}\sqrt{1-\Pi^2}\partial_M
g_{KL}\partial_\alpha x^K
\partial_\beta x^L g^{\beta\alpha} \sqrt{-\det g}+
\nonumber \\
&+&\Pi H_{MKL}\partial_\tau x^K\partial_\sigma
x^L=0 \ , \nonumber \\
\end{eqnarray}
where
\begin{equation}
H_{MNK}=\partial_M b_{NK}+\partial_N b_{KM}+
\partial_K b_{MN} \ .
\end{equation}
To proceed further note that in  case of the WZW model $H_{MNK}$ obeys following
relation
\begin{equation}
H_{MNK}E^{M}_{ \ A}E^N_{ \ B} E^K_{ \ C}= \kappa f_{ABC} \ ,
\end{equation}
where $\kappa$ is a constant. Then we can  write
\begin{equation}
H_{MKL}\partial_\tau x^K \partial_\sigma x^L= \kappa E_M^{ \ A}
f_{ABC}J^B_\tau J^C_\sigma \ .
\end{equation}
Using this result
we finally find that  the equation of motion for $x^M$ has the form
\begin{eqnarray}
\sqrt{1+\Pi^2}K_{AB}\partial_\alpha[J_\beta^B g^{\beta\alpha}
\sqrt{-\det g}] +\Pi\kappa  f_{ABC}J^B_\tau
J^C_\sigma=0 \ .  \nonumber \\
\end{eqnarray}
Let us consider Lax connection  in the form
\begin{eqnarray}\label{flatLb}
L_\tau^A=AJ_\tau^A+B\sqrt{-\det g}g^{\sigma\alpha}J_\alpha^A \ ,
\nonumber \\
L_\sigma^A=AJ_\sigma^A-B\sqrt{-\det g}g^{\tau\alpha}J_\alpha^A \ ,
\nonumber \\
\end{eqnarray}
where  parameters $A,B$ will be determined by the requirement
that the Lax  connection should be flat. Explicitly, we have
\begin{eqnarray}
\partial_\tau L_\sigma^A-\partial_\sigma L_\tau^A
=-AJ_\tau^BJ_\sigma^Cf_{BC}^{ \quad A}+B\frac{\Pi}{\sqrt{1+\Pi^2}}
\kappa
f^{\quad A}_{BC}J^B_\tau J^C_\sigma \nonumber \\
\end{eqnarray}
using the equation of motion for $x^M$ and also $\partial_\tau
J_\sigma^A-\partial_\sigma J_\tau^A+J^C_\tau J^D_\sigma f_{CD}^{
\quad  A}=0$.
 At the same time we
have
\begin{eqnarray}
f^{\quad A}_{BC}L^B_\tau L^C_\sigma=
(A^2-B^2) f^{\quad A}_{BC}J^B_\tau J^C_\sigma \ .   \nonumber \\
\end{eqnarray}
Then the   requirement that the Lax current should be flat leads to
the equation
\begin{equation}
A^2-B^2-A+\frac{\kappa \Pi B}{\sqrt{1+\Pi^2}}=0 \ .
\end{equation}
If we presume the solution in the form $B=-\Lambda A$ we find
\begin{equation}
A=\frac{1}{1-\Lambda^2}\left(1+\Lambda
\kappa\frac{\Pi}{\sqrt{1+\Pi^2}}\right) \ , \quad
B=-\frac{\Lambda}{1-\Lambda^2} \left(1+\Lambda
\kappa\frac{\Pi}{\sqrt{1+\Pi^2}}\right) \ .
\end{equation}
It is important to stress that the values of $A$ and $B$ were
determined on condition when all fields obey the equations of
motion which implies that
\begin{equation}\label{PiF}
\frac{\Pi}{\sqrt{1+\Pi^2}}
=\frac{(2\pi\alpha')F_{\tau\sigma}+b_{\tau\sigma}}{\sqrt{-\det g}} \
.
\end{equation}
Then it is natural to  propose off-shell form of the flat current when we use
(\ref{PiF}) in (\ref{flatLb}) so that
\begin{eqnarray}\label{LWZWgen}
L_\tau^A=\frac{1}{1-\Lambda^2}
\left(1+\Lambda\kappa\frac{(2\pi\alpha')F_{\tau\sigma}+b_{\tau\sigma}}{
\sqrt{-\det g}}\right)(J_\tau^A-\Lambda \sqrt{-\det
g}g^{\sigma\alpha}J_\alpha^A) \ ,
\nonumber \\
L_\sigma^A=\frac{1}{1-\Lambda^2}
\left(1+\Lambda\kappa\frac{(2\pi\alpha')F_{\tau\sigma}+b_{\tau\sigma}}{
\sqrt{-\det g}}\right)(J_\tau^A+\Lambda \sqrt{-\det
g}g^{\tau\alpha}J_\alpha^A) \ .
\nonumber \\
\end{eqnarray}
Clearly this current is flat which is necessary condition for the
integrability of D1-brane theory on the group manifold with non-zero
$b_{NS}$ field.  However we have to also show that the Poisson
bracket between spatial components of the Lax connection
(\ref{LWZWgen}) has the right form in order to ensure an infinite
number of conserved charges in involution.
\section{Hamiltonian Formalism}\label{fifth}
In this section we  calculate the algebra of Poisson brackets of Lax
connection for the model defined in the previous section.
 To do this we have to develop corresponding Hamiltonian formalism.
  From the action (\ref{D1Baction}) we find
\begin{eqnarray}
p_M&=&T_{D1} \frac{1}{\sqrt{-\det g -((2\pi\alpha')F_{\tau\sigma}+
b_{\tau\sigma})^2}}(g_{MN}\partial_\alpha x^N g^{\alpha \tau}\det g+
((2\pi\alpha')F_{\tau\sigma}+b_{\tau\sigma})b_{MN}\partial_\sigma
x^N) \ , \nonumber \\
\pi^\sigma&=&\frac{T_{D1}(2\pi\alpha')((2\pi\alpha')F_{\tau\sigma}+
b_{\tau\sigma})}{\sqrt{-\det g -((2\pi\alpha')F_{\tau\sigma}+
b_{\tau\sigma})^2}} \ , \quad
\pi^\tau\approx 0 \ . \nonumber \\
\end{eqnarray}
Using these definitions we find that    the bare Hamiltonian is
equal to
\begin{eqnarray}
H_E=\int d\sigma (p_M\partial_\tau x^M+\pi^\sigma \partial_\tau
A_\sigma-\mL)= -\int d^2\sigma \pi^\sigma\partial_\sigma A_\tau
\nonumber \\
\end{eqnarray}
while we have two  primary constraints
\begin{eqnarray}
\mH_\tau &\equiv & (p_M-\frac{\pi^\sigma}{(2\pi\alpha')}
b_{MK}\partial_\sigma x^K) g^{MN}
(p_K-\frac{\pi^\sigma}{(2\pi\alpha')}b_{NL}\partial_\sigma
x^L)+\nonumber \\
&+& T_{D1}^2g_{MN}\partial_\sigma x^M\partial_\sigma
x^N+\frac{(\pi^\sigma)^2}{(2\pi\alpha')^2}g_{MN}\partial_\sigma
x^M\partial_\sigma x^N\approx 0 \ ,
\nonumber \\
\mH_\sigma &\equiv &
p_M\partial_\sigma x^M\approx 0 \ . \nonumber \\
\end{eqnarray}
Then the extended Hamiltonian has the form
\begin{equation}
H=\int d\sigma (\lambda_\tau\mH_T+\lambda_\sigma
\mH_\sigma+A_\tau\partial_\sigma\pi^\sigma+v_\tau \pi^\tau) \ ,
\end{equation}
where $\lambda_\tau,\lambda_\sigma,v_\tau$ are Lagrange multipliers
corresponding to the primary constraints.
Further,  the requirement of the preservation of the primary
constraint $\pi^\tau\approx 0$ implies the secondary constraint
\begin{equation}
\mG=\partial_\sigma \pi^\sigma\approx 0 \ .
\end{equation}
Now we proceed to the analysis of the stability of the primary
constraints $\mH_\tau,\mH_\sigma$.
In fact,  after some algebra we find
\begin{eqnarray}
& &\pb{\mH_\tau(\sigma),\mH_\tau(\sigma')}=
 8\left(T_{D1}^2+\left(\frac{\pi^\sigma}{2\pi\alpha'}\right)^2\right)
\mH_\sigma\partial_\sigma \delta(\sigma-\sigma')+
\nonumber \\
&+&4\left(T_{D1}^2+\left(\frac{\pi^\sigma}{2\pi\alpha'}\right)^2
\right)
\partial_\sigma \mH_\sigma \delta(\sigma-\sigma')+
\frac{8}{(2\pi\alpha')^2}\pi^\sigma \mG \mH_\sigma
\delta(\sigma-\sigma')\ . \nonumber \\
\end{eqnarray}
We see that this Poisson brackets vanishes on the constraint surface
$\mH_\sigma\approx 0 \ , \mG\approx 0 $. In the same way we find
\begin{eqnarray}
\pb{\mH_\sigma(\sigma),\mH_\tau(\sigma')}=
\mH_\tau\partial_\sigma \delta(\sigma-\sigma')+
\partial_\sigma \mH_\tau \delta(\sigma-\sigma')+
\frac{1}{(\pi\alpha')^2}\pi^\sigma \partial_\sigma x^M
g_{MN}\partial_\sigma x^N \mG\delta(\sigma-\sigma') \ ,  \nonumber \\
\end{eqnarray}
where again right side of this Poisson bracket vanishes on the
constraint surface. Finally we obtain
\begin{equation}
\pb{\mH_\sigma(\sigma),\mH_\sigma(\sigma')}=2\mH_\sigma(\sigma)
\partial_\sigma \delta(\sigma-\sigma')+\partial_\sigma
\mH_\sigma(\sigma)\delta(\sigma-\sigma')
\end{equation}
and we again see that this Poisson bracket vanishes on the
constraint surface. Collecting all these results we find that the
constraints $\mH_\tau\approx 0 \ , \mH_\sigma\approx 0$ are
preserved during the time evolution of the system and no further
constraints are generated.
Now we are ready to proceed to the calculation of the Poisson
bracket between spatial component of the Lax connection that
expressed using canonical variables has the form
%
\begin{eqnarray}
L_\sigma^A&=&\frac{1}{1-\Lambda^2} \left(1+\Lambda \kappa
\frac{\frac{\pi^\sigma}{2\pi\alpha'}} {\sqrt{T_{D1}^2+
\left(\frac{\pi^\sigma}{2\pi\alpha'}\right)^2}}\right)\left(J^A_\sigma-
\right. \nonumber \\
&-&\left. \Lambda K^{AB}E^M_{  \
B}\frac{1}{\sqrt{T_{D1}^2+\left(\frac{\pi^\sigma}{2\pi\alpha'}\right)^2}}
(p_M-\frac{\pi^\sigma}{2\pi\alpha'}b_{MN}\partial_\sigma
x^N)\right) \ . \nonumber \\
\end{eqnarray}
Then after some calculations we obtain
\begin{eqnarray}
\pb{L_\sigma^A(\Lambda,\sigma), L_\sigma^B(\Gamma,\sigma')}&=&
f(\Lambda)f(\Gamma) f_{FE}^{ \quad  A}K^{EB} K^{FC} E^M_{ \ C}
(p_M-\frac{\pi^\sigma}{2\pi\alpha'}b_{MN}\partial_\sigma x^N)
\delta(\sigma-\sigma') +\nonumber \\
&+&f(\Lambda)f(\Gamma)\frac{\pi^\sigma}{2\pi\alpha'}
 \kappa f_{EF}^{ \quad A}K^{FB}J^E_\sigma\delta(\sigma-\sigma')-
\nonumber \\
&-&(h(\Lambda)f(\Gamma)+f(\Lambda)h(\Gamma))f_{EC}^{ \quad
A}K^{CB}J^E_\sigma \delta(\sigma-\sigma')-\nonumber \\
&-&(h(\Lambda)f(\Gamma)+f(\Lambda)h(\Gamma))K^{AB}
\partial_\sigma \delta(\sigma-\sigma')-\nonumber \\
&-&[h(\Lambda)\frac{d}{d\pi^\sigma}f(\Gamma)+ f(\Lambda)
\frac{d}{d\pi^\sigma}h(\Gamma)]K^{AB}\mG\delta(\sigma-\sigma') \ ,
 \nonumber \\
\end{eqnarray}
where
\begin{eqnarray}
f(\Lambda)&=& \frac{1}{1-\Lambda^2} \left(1+\Lambda \kappa
\frac{\frac{\pi^\sigma}{2\pi\alpha'}} {\sqrt{T_{D1}^2+
\left(\frac{\pi^\sigma}{2\pi\alpha'}\right)^2}}\right)
\frac{\Lambda}{\sqrt{T_{D1}^2+\left(\frac{\pi^\sigma}{2\pi\alpha'}\right)^2}}
\ , \nonumber \\
h(\Gamma)&=& \frac{1}{1-\Lambda^2} \left(1+\Lambda \kappa
\frac{\frac{\pi^\sigma}{2\pi\alpha'}} {\sqrt{T_{D1}^2+
\left(\frac{\pi^\sigma}{2\pi\alpha'}\right)^2}}\right) \ , \nonumber
\\
\end{eqnarray}
and where we used the fact that
\begin{equation}
\partial_\sigma f\equiv \frac{df}{d\pi^\sigma}\mG \ .
\end{equation}
 We again demand that the expression proportional to the delta
function on the right side is equal to
\begin{eqnarray}
-(AL^C_\sigma(\Lambda)-BL_\sigma^C(\Gamma))f_{CD}^{ \quad  A}K^{DB}
\nonumber \\
\end{eqnarray}
so that we find
\begin{eqnarray}
A=\frac{h(\Gamma)}{\sqrt{T_{D1}^2+
(\frac{\pi^\sigma}{2\pi\alpha'})^2}}\frac{\Gamma^2}{\Gamma-\Lambda}
\left[1-\kappa\Lambda\frac{\pi^\sigma}{2\pi\alpha' \sqrt{
T_{D1}^2+ (\frac{\pi^\sigma}{2\pi\alpha'})^2}}\right] \ , \nonumber \\
B=\frac{h(\Lambda)}{\sqrt{T_{D1}^2+
(\frac{\pi^\sigma}{2\pi\alpha'})^2}}\frac{\Lambda^2}{\Gamma-\Lambda}
\left[1-\kappa\Gamma\frac{\pi^\sigma}{2\pi\alpha' \sqrt{
T_{D1}^2+ (\frac{\pi^\sigma}{2\pi\alpha'})^2}}\right] \ . \nonumber \\
\end{eqnarray}
In summary we obtain following form of the Poisson bracket between
spatial component of the Lax connection
\begin{eqnarray}\label{pbLbns}
& &\pb{L_\sigma^A(\Lambda,\sigma), L_\sigma^B(\Gamma,\sigma')}=
\nonumber \\
&=& -\frac{h(\Gamma)}{\sqrt{T_{D1}^2+
(\frac{\pi^\sigma}{2\pi\alpha'})^2}}\frac{\Gamma^2}{\Gamma-\Lambda}
\left[1-\kappa\Lambda\frac{\pi^\sigma}{2\pi\alpha' \sqrt{ T_{D1}^2+
(\frac{\pi^\sigma}{2\pi\alpha'})^2}}\right]L^C_\sigma(\Lambda)
f_{CD}^{ \quad  A}K^{DB}\delta(\sigma-\sigma')+\nonumber \\
&+&\frac{h(\Lambda)}{\sqrt{T_{D1}^2+
(\frac{\pi^\sigma}{2\pi\alpha'})^2}}\frac{\Lambda^2}{\Gamma-\Lambda}
\left[1-\kappa\Gamma\frac{\pi^\sigma}{2\pi\alpha' \sqrt{ T_{D1}^2+
(\frac{\pi^\sigma}{2\pi\alpha'})^2}}\right]L_\sigma^C(\Gamma)f_{CD}^{\quad
 A}K^{DB}\delta(\sigma-\sigma') -\nonumber \\
&-&\left(h(\Lambda)f(\Gamma)+f(\Lambda)h(\Gamma)\right)K^{AB}
\partial_\sigma \delta(\sigma-\sigma')-\nonumber \\
&-&\left(h(\Lambda)\frac{d}{d\pi^\sigma}f(\Gamma)+ f(\Lambda)
\frac{d}{d\pi^\sigma}h(\Gamma)\right)K^{AB}\mG\delta(\sigma-\sigma')
\ .
 \nonumber \\
\end{eqnarray}
We see that the right side of this Poisson bracket has similar  form
(up to terms proportional to the primary constraints $\mG$) as in
case of WZW model. More explicitly, we find that it depends on the
expression $\frac{\pi}{2\pi\alpha'}$ which is constant on-shell and
determines the contribution of the fundamental strings to the
resulting tension of the bound state of D1-brane and fundamental
string and also to the coupling with $b_{NS}$ two form. However the
fact that the Poisson bracket between spatial components of the Lax
connection takes the standard form implies that  D1-brane on the
group manifold with non-trivial NS-NS two form is integrable and
possesses infinite number of conserved charges in involution, after
appropriate regularization of the Poisson bracket (\ref{pbLbns}).

\subsection*{Acknowledgement}
 This work was supported by the Grant
Agency of the Czech Republic under the grant P201/12/G028.


\begin{thebibliography}{20}
\bibitem{Bena:2003wd}
  I.~Bena, J.~Polchinski and R.~Roiban,
\emph{``Hidden symmetries
 of the $AdS(5) x S**5$  superstring,''}
  Phys.\ Rev.\ D {\bf 69} (2004) 046002
  [hep-th/0305116].

\bibitem{Sfondrini:2014via}
  A.~Sfondrini,
\emph{``Towards integrability for AdS3/CFT2,''}
  arXiv:1406.2971 [hep-th].

\bibitem{Puletti:2010ge}
  V.~G.~M.~Puletti,
\emph{``On string integrability: A Journey through the
two-dimensional hidden symmetries in the AdS/CFT dualities,''}
  Adv.\ High Energy Phys.\  {\bf 2010} (2010) 471238
  [arXiv:1006.3494 [hep-th]].


\bibitem{vanTongeren:2013gva}
  S.~J.~van Tongeren,
\emph{``Integrability of the $AdS_5 \times S^5$ superstring and its
deformations,''}
  arXiv:1310.4854 [hep-th].

\bibitem{Arutyunov:2009ga}
  G.~Arutyunov and S.~Frolov,
 \emph{"Foundations of the $AdS_5 x S^5$ Superstring. Part I,''}
  J.\ Phys.\ A {\bf 42} (2009) 254003
  [arXiv:0901.4937 [hep-th]].

\bibitem{Dorey:2006mx}
  N.~Dorey and B.~Vicedo,
\emph{``A Symplectic Structure for String Theory on Integrable
Backgrounds,''}
  JHEP {\bf 0703} (2007) 045
  [hep-th/0606287].




\bibitem{Maillet:1985ek}
  J.~M.~Maillet,
\emph{``New Integrable Canonical Structures in Two-dimensional
Models,''}
  Nucl.\ Phys.\ B {\bf 269} (1986) 54.

\bibitem{Maillet:1985ec}
  J.~M.~Maillet,
\emph{``Hamiltonian Structures for Integrable Classical Theories
From Graded Kac-moody Algebras,''}
  Phys.\ Lett.\ B {\bf 167} (1986) 401.


\bibitem{Delduc:2012qb}
  F.~Delduc, M.~Magro and B.~Vicedo,
\emph{``Alleviating the non-ultralocality of coset sigma models
through a generalized Faddeev-Reshetikhin procedure,''}
  JHEP {\bf 1208} (2012) 019
  [arXiv:1204.0766 [hep-th]].

\bibitem{Delduc:2012vq}
  F.~Delduc, M.~Magro and B.~Vicedo,
\emph{``Alleviating the non-ultralocality of the $AdS_5 x S^5$
superstring,''}
  JHEP {\bf 1210} (2012) 061
  [arXiv:1206.6050 [hep-th]].

\bibitem{Myers:1999ps}
  R.~C.~Myers,
 \emph{``Dielectric branes,''}
  JHEP {\bf 9912} (1999) 022
  [hep-th/9910053].

\end{thebibliography}
\end{document}